\documentclass[aps,showpacs,twocolumn]{revtex4}
\usepackage{amsfonts}
\usepackage{amsmath}
\usepackage{amssymb}
\usepackage{graphicx}
\usepackage{bm}

\input{tcilatex}
\begin{document}

\title{Thermodiffusion in binary liquids: The role of irreversibility}
\author{Alois W\"{u}rger}
\affiliation{LOMA, Universit\'{e} de Bordeaux \& CNRS, 351 cours de la Lib\'{e}ration,
33405 Talence, France}

\begin{abstract}
We study thermal diffusion in binary mixtures in the framework of
non-equilibrium thermodynamics. Our formal result displays the role of
partial enthalpies $h_{i}$ and Onsager's generalized mobilities $A_{i}$. The
ratio $A_{1}/A_{2}$ provides a measure for the irreversible character of
thermal diffusion. Comparison with experimental data on benzene,
cyclohexane, toluene, and $n$-alkanes shows that irreversibility is
essential for thermal diffusion, and in particular for the isotope effect. \ 

\ 

PACS\ numbers 66.10.C-; 05.70.Ln; 82.70.-y
\end{abstract}

\maketitle

\section{Introduction}

Thermal diffusion, or the Soret effect, describes the mass flow induced by a
temperature gradient in complex fluids \cite{Wie04,Pia08,Wue10}. Together
with the Dufour effect that acccounts for heat flow in a concentration
gradient, it constitutes a classical example of Onsager's reciprocal
relations for non-equilibrium systems \cite{deG84}. Thermal diffusion is one
of the mechanisms governing the compositional grading in the Earth's
petroleum reservoirs \cite{Gal09} and the isotope fractionation in silicate
melts \cite{Dom11}. In recent years, thermal diffusion has proven a
versatile means for confining colloidal suspensions \cite{Wie10,Mae11},
manipulating DNA\ through nanopores \cite{He13}, or self-propelling Janus
particles \cite{Jia10,Qia13}.

The Soret effect of macromolecules and colloidal particles is to a large
extent determined by viscous effects and thus can be treated in the
framework of macroscopic hydrodynamics \cite{Wue10}. Exploiting the
reciprocal laws for heat and mass flows, Derjaguin related thermally driven
transport to the solute-solvent interaction enthalpy \cite{Der87}. Much
recently much progress has been made for charged solute, confirming
Derjaguin's picture of enthalpy flow as driving mechanism and pointing the
out the role of the electrolyte Seebeck effect \cite{Put05,Wue08,Vig10}.

Much less is known, however, on thermal diffusion in molecular mixtures. In
spite of the many experimental \cite%
{Koe95,Zha96,Deb01,Kit04,Pol06,Lea07,Bla08,Mad10} and molecular dynamics
data \cite{Gal03,Art07,Art08} on the composition, temperature, and mass
dependencies, available models diverge on the drivinng mechanism \cite%
{Haa49, Kem89,Shu98,Wue09,Esl09,Vil11,Wue13,Har13}. Although the observed
isotope effect \cite{Deb01} indicates irreversible behavior, there is at
present no general agreement whether the Soret effect can be described by
equilibrium thermodynamics, or on the contrary reflects the irreversible
character of the underlying diffusion process.

The present Letter discusses these questions on the basis of formally exact
expressions derived by de Groot \& Mazur half a century ago, yet which so
far have been given no attention. As a main purpose, we separate equilibrium
and irreversible factors of the Soret coefficient. Comparison with recent
experimental findings for binary mixtures, provides a simple physical
picture for the driving mechanism and a clear signature of irreversibility.

Consider a two-component molecular liquid with volume fractions $c_{1}$ and $%
c_{2}=1-c_{1}$. Non-uniform composition and temperature induce a current 
\begin{equation}
J_{1}=-D\nabla c_{1}-c_{1}c_{2}D_{T}\nabla T.  \label{1}
\end{equation}%
The first term or gradient diffusion tends to smoothen composition
inhomogeneities, whereas the second one, or thermal diffusion, pushes one
component to the cold and the other to the hot, and thus favors separation.
The steady state $J_{1}=0$ is characterized by a spatial composition
gradient $\nabla c_{1}=-c_{1}c_{2}S_{T}\nabla T$, where $S_{T}=D_{T}/D$ is
the Soret coefficient. Even for simple systems such as benzene-cyclohexane,
both its sign and its dependence on composition and temperature lack a
rationale so far.

\section{Non-equlibrium thermodynamics}

The coefficients $D$ and $D_{T}$ are related to the thermodynamic forces
introduced by Onsager and worked out in detail by de Groot and Mazur. As a
generalized state function, the Planck potential $\mu _{i}/T$ of component $%
i $ is given by the ratio of its chemical potential $\mu _{i}$ (or partial
Gibbs energy) and the absolute temperature $T$. Just like the gradient of
gravitational or electric potentials give mechanical forces, that of the
Planck potential defines thermodynamic forces acting on each molecular
species. For an otherwise homogeneous system one has 
\begin{equation}
\nabla \frac{\hat{\mu}_{i}}{T}=-\frac{\hat{h}_{i}}{T^{2}}\nabla T+\frac{\hat{%
\mu}_{ii}^{c}}{T}\nabla c_{i},  \label{4}
\end{equation}%
where the first term arises from the Gibbs-Helmholtz relation $d(\hat{\mu}%
_{i}/T)/dT=-\hat{h}_{i}/T^{2}$ and the second one from applying the
Gibbs-Duhem relation to $\hat{\mu}_{ij}^{c}=d\hat{\mu}_{i}/dc_{j}$. For
incompressible liquids, it is convenient to use volume specific quantities,
e.g., $\hat{\mu}_{i}=\mu _{i}/v_{i}$ the volume per molecule $v_{i}$; thus $%
\hat{h}_{i}$ is the corresponding enthalpy density.

The thermodynamic force on species $j$ induces a current $a_{ij}\nabla (\hat{%
\mu}_{j}/T)$ of species $i$, where the mobility matrix $a_{ij}$ is symmetric
and positive definite. Eliminating the related heat flow and accounting for $%
J_{2}=-J_{1}$ results in \cite{deG84} 
\begin{equation}
J_{1}=-\left( c_{2}a_{11}-c_{1}a_{12}\right) \nabla \frac{\hat{\mu}_{1}}{T}%
+\left( c_{1}a_{22}-c_{2}a_{21}\right) \nabla \frac{\hat{\mu}_{2}}{T}.
\label{2}
\end{equation}%
Inserting (\ref{4}) and separating thermal and concentration gradients, one
readily identifies $D$ and $D_{T}$\ in (\ref{1}), \ 

\begin{equation}
D=\frac{c_{1}(c_{1}a_{22}-c_{2}a_{12})+c_{2}(c_{2}a_{11}-c_{1}a_{12})}{%
c_{1}c_{2}T}c_{1}\hat{\mu}_{11}^{c},  \label{10}
\end{equation}

\begin{equation}
D_{T}=\frac{(c_{1}a_{22}-c_{2}a_{12})\hat{h}_{2}-(c_{2}a_{11}-c_{1}a_{12})%
\hat{h}_{1}}{c_{1}c_{2}T^{2}}.  \label{12}
\end{equation}%
These relations for the diffusion and thermodiffusion coefficients have been
obtained by de Groot and Mazur, albeit for mass instead of volume fractions 
\cite{deGroot}. They reveal several remarkable features. First, the
coefficient $D_{T}$\ is proportional to the partial enthalpies of the
components; this confirms the central role of enthalpy for thermo-osmosis
pointed out by Derjaguin \cite{Der87}.\ 

Second, the factors $a_{ij}$ accounting for irreversibility, appear in both $%
D$ and $D_{T}$ in the form $B_{2}=c_{1}a_{22}-c_{2}a_{12}$ and $%
B_{1}=c_{2}a_{11}-c_{1}a_{12}$. In view of a two-paramter model discussed
below, we absorb the chemical potential derivative in these coefficients and
separate the dynamic viscosity $\eta $. Thus we put $A_{i}=B_{i}(\eta
/c_{1}c_{2}T)c_{1}\hat{\mu}_{11}$ and have for the diffusion coefficient 
\begin{equation}
D=\frac{c_{1}A_{2}+c_{2}A_{1}}{\eta }.  \label{14}
\end{equation}%
Note thato in general both $a_{ij}$ and $A_{i}$ are functions of composition
and temperature.\ Inserting the $A_{i}$ in the thermal diffusion coefficient
we have 
\begin{equation}
D_{T}=\frac{1}{c_{1}\hat{\mu}_{11}^{c}\eta T}\left( A_{2}\frac{h_{2}}{v_{2}}%
-A_{1}\frac{h_{1}}{v_{1}}\right) .  \label{16}
\end{equation}%
The derivative of the chemical potential\ can be split in two factors,%
\begin{equation}
c_{1}\hat{\mu}_{11}^{c}=\frac{k_{B}T}{c_{1}v_{2}+c_{2}v_{1}}\Gamma ,
\label{18}
\end{equation}%
where the first one arises from the mixing entropy; the remaining factor\ $%
\Gamma $ is related to composition dependence of the \textquotedblleft
activity coefficient\textquotedblright\ $\gamma $ and thus accounts for
non-ideal behavior \cite{deG84}.

The remainder of this paper deals with the Soret coefficient $S_{T}=D_{T}/D$%
. For its derivation it was advantageous to consider volume fractions $%
c_{i}=n_{i}v_{i}$, where $n_{i}$ is the concentration. Since Soret data and
thermodynamic excess quantities of binary mixtures are usually given in
terms of mole fractions $x_{i}=n_{i}/(n_{1}+n_{2})$, we change variables and
obtain 
\begin{equation}
S_{T}=\frac{1}{\Gamma k_{B}T^{2}}\frac{h_{2}A_{2}/v_{2}-h_{1}A_{1}/v_{1}}{%
x_{1}A_{2}/v_{2}+x_{2}A_{1}/v_{1}}.  \label{19}
\end{equation}%
This formally exact expression reveals the intricate equilibrium and
irreversibiltiy aspects of $S_{T}$: The state function enthalpy is an
equilibrium property, whereas the mobilities $A_{i}$ account for the
irreversible nature of thermal diffusion.

It turns out instructive to consider the case where the mobility
coefficients are constant and identical to each other. Then the Soret
coefficient takes the form 
\begin{equation}
S_{T}=\frac{1}{\Gamma k_{B}T^{2}}\frac{h_{2}/v_{2}-h_{1}/v_{1}}{%
x_{1}/v_{2}+x_{2}/v_{1}}\ \ \ \ \ \ \ (A_{1}=A_{2}),  \label{13}
\end{equation}%
which is proportional to the difference in partial enthalpy per volume. In
other words, the molecules with the more negative enthalpy density migrates
to the cold, similar to gravity-driven sedimentation where the denser
component accumulate at the bottom. Remarkably, the mobilities have dropped
out, such that $S_{T}$ depends on equilibrium properties only. This
expression was first obtained by Haase for gas mixtures, with mole fractions
weighted by molecular mass instead of volume \cite{Haa49}; later on it has
been refined by several authors \cite{Kem89,Shu98}.

Yet the equilibrium hypothesis is not generally valid, and in particular
fails for molecular liquids. With the enthalpy and volume parameters given
in the Tables, Haase's expression (\ref{13})\ results in a strong positive
Soret coefficient for benzene in cyclohexane, whereas the data \cite{Deb01}
plotted in Fig. 1 rather show the opposite behavior. Even more strikingly,
it misses the isotope effect: Since protonated and deuterated benzene, C$%
_{6} $H$_{6}$ and C$_{6}$D$_{6}$, hardly differ in their enthalpies and
molecular volumes, (\ref{13}) gives the same Soret behavior, whereas the
measured values are higher for the heavier isotope.

\begin{figure}[b]
\includegraphics[width=\columnwidth]{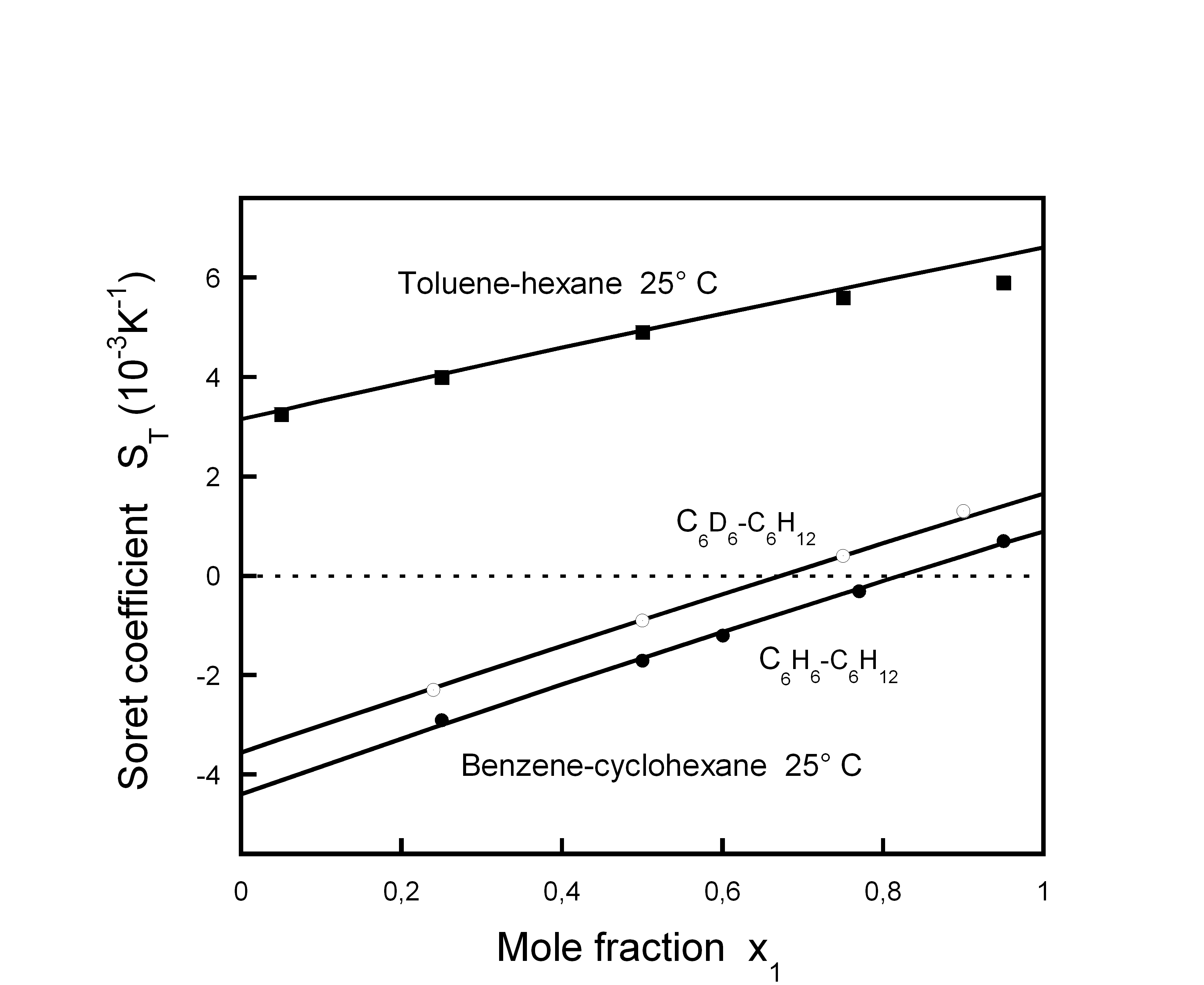}
\caption{Composition dependence of the Soret coefficient. The
data for benzene-cyclohexane are from Debuschewitz and K\"{o}hler 
\protect\cite{Deb01}, those for toluene-hexane from Zhang et al. 
\protect\cite{Zha96}. The solid lines are calculated from Eq. (\protect\ref%
{22}) with parameters from the tables. For protenated benzene we have used $%
A_{1}/A_{2}=0.765$, and for deuterated benzene $A_{1}/A_{2}=0.785$. }
\end{figure}

\section{The role of enthalpy}

The most important parameters in (\ref{19}) are the partial enthalpies $%
h_{i} $ and the mobility factors $A_{i}$. The former, which are equilibrium
properties, express the fact that thermal diffusion is driven by enthalpy
flow, as is obvious from the forces (\ref{4}), whereas the latter account
for the irreversible character of thermal diffusion.

Equilibrium thermodynamics is mainly concerned with enthalpy differences
with respect to some arbitrary reference state. Eq. (\ref{19}) on the
contrary requires \textit{absolute} values, which are neither easily
measured nor easily calculated. The quantities $h_{i}$ give the enthalpy
change upon removing one molecule from the liquid; they are essentially
determined by molecular interactions; typical values correspond to ten times
the thermal energy and by far exceed conformational and volume expansion
contributions.

The partial enthalpies $h_{i}$ in general depend on composition. In the
following we consider \textquotedblleft regular mixtures\textquotedblright\
where the mean molar value is a parabolic function of the mole fractions, $%
h=x_{1}h_{1}^{0}+x_{2}h_{2}^{0}+x_{1}x_{2}h_{E}$, with the pure-body
enthalpies of molecular interactions$\ h_{i}^{0}$ and the change upon mixing 
$h_{E}$. Then the partial enthalpies take the form 
\begin{equation}
h_{1}=h_{1}^{0}+x_{2}h_{E},\ \ \ h_{2}=h_{2}^{0}+x_{1}h_{E}.  \label{20}
\end{equation}%
Most molecular liquids have positive excess enthalpy of the order of a few
kJ/mole, with the noticeable exception of alcohols in water \cite{DDB}. In
the following we identify $\ h_{i}^{0}$ with the vaporization enthalpies
given in Table I. 
\begin{table}[tbp]
\caption{Pure-component parameters at 25$%
{{}^\circ}%
$ C: vaporization enthalpy $h_{\text{vap}}$; molar volume $v$; viscosity $%
\protect\eta $; specific heat at normal pressure \protect\cite{DDB,NIST}.
Throughout this paper we identify $h_{\text{vap}}$ with the interaction
enthalpy $h^{0}$.\ }%
\begin{tabular}{|l|c|c|c|c|}
\hline
& $%
\begin{array}{c}
h_{\text{vap}}\equiv h^{0} \\ 
\text{(kJ/mol)}%
\end{array}%
$ & $%
\begin{array}{c}
v \\ 
\text{(cm}^{3}\text{/mol)}%
\end{array}%
$\  & $%
\begin{array}{c}
\eta \\ 
\text{(mPa.s)}%
\end{array}%
$ & $%
\begin{array}{c}
C_{P} \\ 
\text{(J/molK)}%
\end{array}%
$ \\ \hline
benzene & $-33.9$ & $89$ & $0.61$ & $135$ \\ \hline
cyclohexane & $-33.3$ & $108$ & $0.88$ & $156$ \\ \hline
toluene & $-38$ & $106$ & $0.59$ & $156$ \\ \hline
hexane & $-30.7$ & $132$ & $0.31$ & $198$ \\ \hline
heptane & $-35.4$ & $147$ & $0.39$ & $225$ \\ \hline
\end{tabular}%
\end{table}

The partial molar volumes $v_{i}$ change upon mixing and have been discussed
as one source for the composition dependence of $S_{T}$ \cite{Har13}. For
the systems considered here, the excess volume $v_{E}$ is rather small and
thus has been neglected from the beginning. For example, for
benzene-cyclohexane mixtures one has $v_{E}=0.65$\ cm$^{3}$/mol, which
corresponds to less than one percent of the molar volumes \cite{DDB},
whereas the excess enthalpy $h_{E}$ amounts to about 10 \% of $h_{i}^{0}$.

The thermodynamic factor $\Gamma $ defined in (\ref{18}) describes the
non-ideal behavior of the chemical potential. In analogy to the enthalpy,
the excess chemical potential $\mu _{E}$ is exptected to be much smaller
than the ideal values $\mu _{i}^{0}$.\ This implies that the activity
coefficient $\gamma $ \ and the quantity $\Gamma $ vary only weakly with
composition. Since the latter appears as an overall factor in $S_{T}$, its
modulation is of little relevance and we may take $\Gamma $ as a constant.
On the contrary, the ideal enthalpies in the numerator of (\ref{19}) largely
cancel each other, thus enhancing the weight of the excess $h_{E}$.

The mobilities $A_{i}$ account for the irreversible character of thermal
diffusion. There is no thermodynamic theory for calculating these
quantities; they have to be extracted from molecular dynamics simulations or
experiments, or to rely on model assumptions. Here we adopt the simplest
two-parameter model that consists in taking $A_{1}$ and $A_{2}$ as
constants. The experimental\ values of Table II are determined from data for
the mutual diffusion coefficients at low dilution, e.g., $A_{1}=(\eta
_{2}/T)D_{1}$ for $x_{1}\rightarrow 0$.

\begin{figure}[b]
\includegraphics[width=\columnwidth]{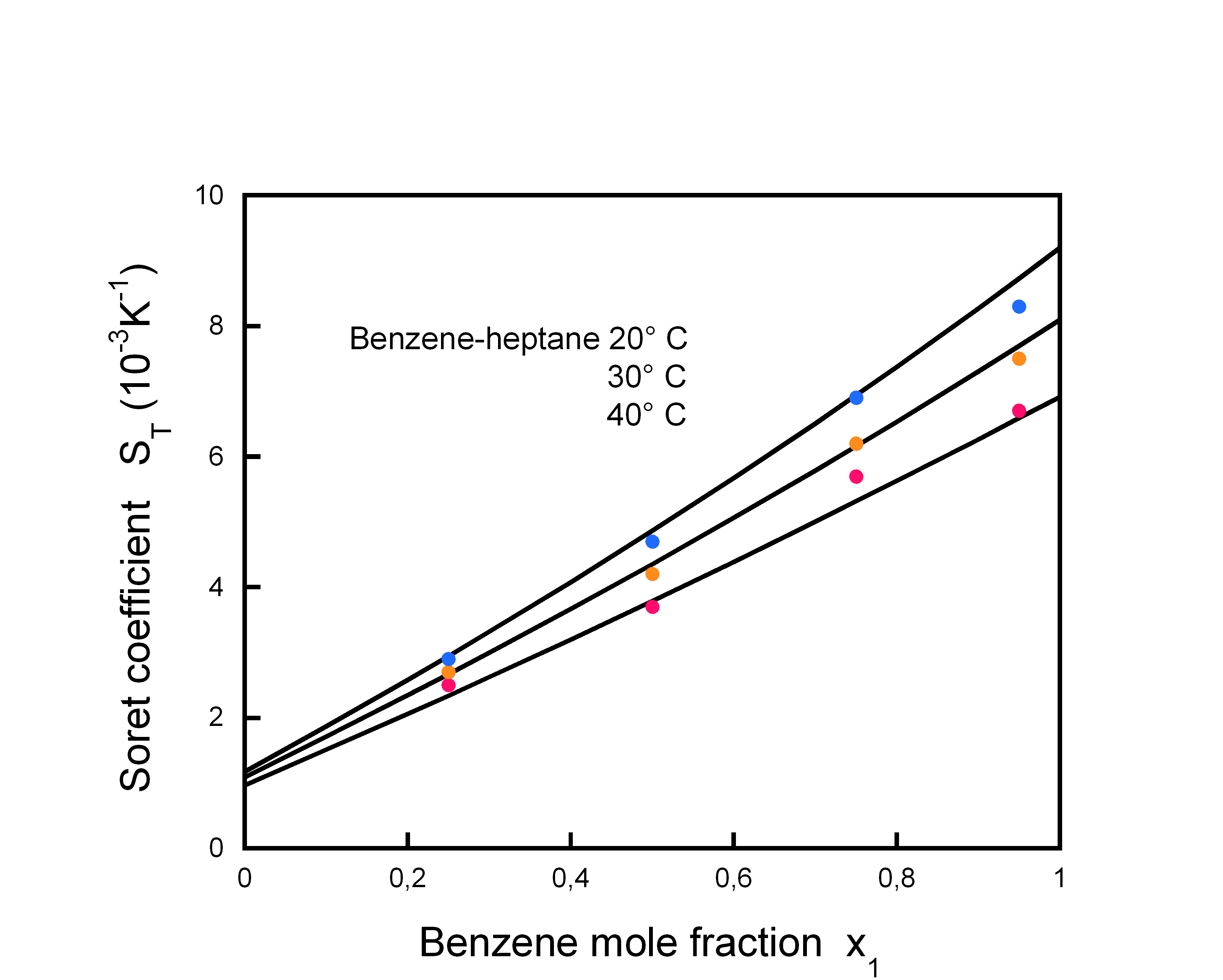}
\caption{Soret coefficient of
benzene-heptane mixtures at three temperatures. Data are from Polyakov et
al. \protect\cite{Pol06}. The solid lines are calculated from (\protect\ref%
{22}) with parameters given in the Tables. }
\end{figure}

\section{Comparison with experiment}

Inserting the above partial enthalpies and defining $\xi =A_{1}/A_{2}$, we
obtain \ 
\begin{equation}
S_{T}=\frac{h_{2}^{0}/v_{2}-\xi h_{1}^{0}/v_{1}+h_{E}(x_{1}/v_{2}-\xi
x_{2}/v_{1})}{\Gamma k_{B}T^{2}(x_{1}/v_{2}+\xi x_{2}/v_{1})}.  \label{22}
\end{equation}%
In Fig. 1 we compare this expression with data for toluene-hexane \cite%
{Zha96} and benzene-cyclohexane \cite{Deb01}, which show an almost linear
variation with the mole fraction of the first component. The theoretical
curves are calculated with the parameters given in the Tables; the fit
values for the irreversibility parameter $A_{1}/A_{2}$ and the excess
enthalpy $h_{E}$ agree rather well with diffusion and thermometry data. Both
the numerator and the denominator of (\ref{22}) contribute equally to the
composition dependence.

Haase's expression (\ref{13}) would result in positive and much too large
value of $S_{T}$. On the other hand, when including the irreversibiltiy
parameter Eq. (\ref{22}) provides a good description for the data.\ The
experimental and fit values of $A_{1}/A_{2}$ agree rather well, yet one
should be aware of the uncertainties of the diffusion coefficients $D_{i}$.
As an important result, these fits show that the Onsager mobilities take a
ratio which significantly differs from unity. This leads us to the
conclusion that irreversibility is crucial for thermal diffusion.

This statement is confirmed by the isotope effect. Since the enthalpy
density is insensitive to the molecular mass, the equilibrium expression (%
\ref{13}) does not differentiate between protonated and deuterated benzene.
On the other hand, the kinetics and thus the mobilities $A_{i}$ depend on
the molecular mass; possible mechanisms are the collision rate as discussed
in a hard-sphere model \cite{Vil11} and the jump rate of activated diffusion 
\cite{Dom11}. Finally we emphasize that the slight change from $%
A_{1}/A_{2}=0.765$ for protonated benzene to $0.785$ for the deuterated
species results in a composition independent offset of $S_{T}$, in good
agreement with the data. 
\begin{table}[tbp]
\caption{Parameters of protonated benzene (B), cyclohexane (CH), toluene
(T), hexane (C$_{6}$), and heptane (C$_{7}$). The coefficient $D_{i}$
describes tracer diffusion of component $i$ \protect\cite{Fun94}. The
parameters $A_{i}$ for tracer diffusison are calculated with the pure
solvent viscosity, e.g. $A_{1}=\protect\eta _{2}D_{1}/T$. Since $S_{T}$
depends on the ratio $A_{1}/A_{2}$ only, we don't give absolute values. The
viscosity of heptane at 40$%
{{}^\circ}%
$ C is 0.33 cP. }%
\begin{tabular}{|l|c|c|c|c|c|c|c|}
\hline
& $D_{1}$ & $D_{2}$ & $A_{1}/A_{2}$ & $A_{1}/A_{2}$ & \multicolumn{2}{|c}{$%
h_{E}\ \text{(kJ/mol)}$} & $\Gamma $ \\ 
& \multicolumn{2}{|c|}{$\text{(10}^{-9}\text{m}^{2}\text{/s) }$} & (exp) & $%
\text{(fit)}$ & (exp) & (fit) &  \\ \hline
B-CH 25$%
{{}^\circ}%
$ & $1.90^{a}$ & $2.09^{a}$ & $0.76$ & $0.765$ & $3.2^{d}$ & $2.8$ & $1.4$
\\ \hline
T-C$_{6}$ 25$%
{{}^\circ}%
$ & $4.0^{b}$ & $2.4^{b}$ & $0.91$ & $0.70$ &  & $1.0$ & $0.7^{f}$ \\ \hline
B-C$_{7}$%
\begin{tabular}{l}
$20%
{{}^\circ}%
$ \\ 
$30%
{{}^\circ}%
$ \\ 
$40%
{{}^\circ}%
$%
\end{tabular}
& $%
\begin{array}{c}
3.92^{c} \\ 
\\ 
4.74^{c}%
\end{array}%
$ & $%
\begin{array}{c}
1.79^{c} \\ 
\\ 
2.28^{c}%
\end{array}%
$ & $%
\begin{array}{c}
0.71^{c} \\ 
\\ 
0.73^{c}%
\end{array}%
$ & 
\begin{tabular}{l}
$0.71$ \\ 
$0.695$ \\ 
$0.67$%
\end{tabular}
& $%
\begin{array}{c}
3.2^{e} \\ 
\\ 
2.5^{e}%
\end{array}%
$ & 
\begin{tabular}{l}
$2.6$ \\ 
$2.3$ \\ 
$2.0$%
\end{tabular}
& 
\begin{tabular}{l}
$1$ \\ 
$1$ \\ 
$1$%
\end{tabular}
\\ \hline
\end{tabular}
\ \ $^{a}$\cite{Fun94};$\ ^{b}$\cite{Koe95};\ $^{c}$Data at 25 and $40%
{{}^\circ}%
$ C \cite{San71}; $^{d}$\cite{DDB}; $^{e}$Data for B-C$_{6}$ at 25 and 40$%
{{}^\circ}%
$ C \cite{DDB}; $^{f}$\cite{Har13}.
\end{table}

Fig. 2 shows the temperature dependence of Soret data reported by Polyakov
et al. for benzene-heptane \cite{Pol06}. An increase of 20 K reduces the
Soret coefficient by one quarter, which corresponds to a relative change of
about 1.2\ \% per K. Comparison with the thermal expansion coefficient $d\ln
v/dT\sim 10^{-3}$ K$^{-1}$, suggests that the change of the molecular volume
is of little relevance. The temperature derivative of the molar enthalpy $%
h^{0}$ is determined by the heat capacities $C_{P}$; with the numbers of
Table I one finds that a relative change of about\ $0.005$ K$^{-1}$, which
accounts for almost half of the variation of the Soret coefficient. The fit
curves in Fig. 2 are obtained by varying in addition both $A_{1}/A_{2}$ and $%
h_{E}$, suggesting that the temperature dependence of $S_{T}$ has not a
single cause.

In the present work we have discarded viscous effects. This is appropriate
for molecular components of similar size, where the two terms of the
thermodynamic force (\ref{4}) carry the same mobilities. On the contrary,
the motion of large particles or polymers is accompanied by viscous flow;
the Einstein coefficient $D$ carries a size-dependent friction factor,
whereas that the thermal diffusion coefficient $D_{T}$\ is constant \cite%
{Wue13}. Mixtures of normal alkanes would provide a model system for the
emergence of viscous effects, because of their rather simple mixing
properties and the available Soret data \cite{Mad10}.

Finally we briefly compare with thermal diffusion in gases. The enthalpy of
an ideal gas, $h=\frac{5}{2}k_{B}T$, is the same for all atoms and
independent of molecular mass and volume. This implies that the Soret
coefficient vanishes if $A_{1}=A_{2}$; in other words, thermal diffusion in
gases and in particular the isotope effect, is a purely non-equilibrium
property. One cannot exclude that the enthalpy of liquid mixtures slightly
depend on the molecular mass, and thus partly accounts for the isotope
effect; yet this possibility is ont supported by the experiments discussed
above.

\section{Conclusion}

In summary, our comparison with experiments provides strong evidence for the
irreversible character of thermal diffusion, even in systems where viscous
effects are absent. Note that the definition of the mobility factors $%
A_{i}=B_{i}(\eta /c_{1}c_{2}T)c_{1}\hat{\mu}_{11}$ in (\ref{14}) and (\ref%
{16}) is not the only and not necessarily the most appropriate choice; in
particular one could prefer to discard the viscosity and activity factors $%
\eta $ and $c_{1}\hat{\mu}_{11}$. The simple two-parameter model used here
could be improved by considering three mobilities $A_{ij}\sim
a_{ij}/c_{i}c_{j}$. For systems where both diffusion and thermal diffusion
data are available, it would be interesting to adjust the coefficients $D$
and $D_{T}$\ independently.

We conclude with a remark on alcohol-water sysems that cannot be described
as regular mixtures. In accordance with (\ref{22}), the measured Soret
coefficient \cite{Kit04} roughly follows the excess enthalpy as a function
of composition \cite{DDB}, and in particular reflects the cusp at $x_{\text{%
eth}}=0.15$; because of the strongly irregular behavior, a more rigorous
comparison would require to go beyond the linear law (\ref{20}) for the
partial enthalpies.

\textbf{Acknowledgment.} The author acknowledges support through the Leibniz
program of Universit\"{a}t Leipzig during the summer term 2013, and thanks
the groups of Frank Cichos and Klaus Kroy for their kind hospitality.


\begin{thebibliography}{99}
\bibitem{deGroot} This formal part of our paper parallels Sect. XI.8 of Ref. 
\cite{deG84}. Rewritten in terms of mass fractions, our Eqs. (\ref{2},\ref%
{10},\ref{12}) correspond to de Groot and Mazur's (281,368,369).
\end{thebibliography}
\end{document}